\documentclass[a4paper,11pt]{article}
\usepackage{pos}
\usepackage[numbers]{natbib}

\title{Asteroid Occultations with VERITAS: Observations and Considerations}
\ShortTitle{Asteroid Occultations with VERITAS}

\author*[a]{Joshua Thomas Bartkoske}
\author*[a]{Anne Duerr}
\author[a]{Dave Kieda}
\onbehalf{on behalf of the VERITAS collaboration}

\affiliation[a]{University of Utah,\\
  Salt Lake City, USA}

\emailAdd{joshua.bartkoske@utah.edu}
\emailAdd{a.duerr@utah.edu}
\emailAdd{dave.kieda@utah.edu}

\abstract{Occultations, the covering up of one celestial body by another celestial body, have been used in astronomy for millennia to learn about the sun and moon. Since 2018, VERITAS has implemented a program to detect predicted asteroid occultations, where an asteroid covers up a star. VERITAS has attempted to observe over 100 occultations to date and successfully observed 20 occultations. With these occultations, VERITAS can directly measure the smallest angular diameters of any instrument or technique in the optical for stars between magnitude 9 and 13. Each angular diameter is measured by fitting the diffraction pattern observed by the central VERITAS pixel at the start and end of an occultation. Once a planned FADC upgrade is complete, VERITAS will begin a program to search for serendipitous occultations within its full field of view (3 deg). Serendipitous occultations of sub-km trans-Neptunian objects (TNOs) have the potential to constrain models of solar system formation. This work details how VERITAS predicts and observes occultations as well as the overall status of the asteroid occultation program and future steps for observing occultations of both asteroids and TNOs.
}

\ConferenceLogo{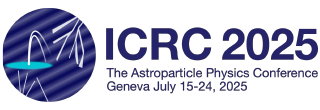}

\FullConference{39th International Cosmic Ray Conference (ICRC2025)\\
 15–24 July 2025\\
Geneva, Switzerland\\}

\begin{document}
\maketitle

\section{Introduction}

Occultations have been used to measure the angular diameters of stars since the first measurement of Regulus in 1936 \citep{1936AcaDeSci_Arnulf_firstLOmeasurement}. Since then, lunar occultations (the moon occulting stars) have been used to measure the angular diameters of hundreds of stars down to ~1 milliarcsecond (mas) \citep{2019ICRC_HassanDaniel_AsO_VERITAS_Overview}. However, most stars have angular diameters below 1 mas. By nature of their increased distance from the observer, asteroid occultations can push angular diameter measurements down to sub-mas values using the same techniques as lunar occultations. Asteroid occultations have been used to directly measure the smallest stellar angular diameters of any instrument in optical wavelengths as demonstrated by the Very Energetic Radiation Imaging Telescope Array System (VERITAS) \citep{2019NatAs_BenbowVERITAS_AsO}. For additional details on the background of asteroid occultations and comparison with other sub-mas methods see \citep{2019ICRC_HassanDaniel_AsO_VERITAS_Overview}.

VERITAS is an array of four imaging atmospheric Cherenkov telescopes (IACTs) in southern Arizona located at the Fred Lawrence Whipple Observatory (FLWO). By nature of being focused on capturing the faint blue Cherenkov radiation in the atmosphere, IACTs are capable of unfiltered optical observations with maximum sensitivity in approximately the B-band. Since 2018, the asteroid occultation program at VERITAS has observed more than 20 successful occultations of stars ranging from magnitude 9.9 to 13.3 in the V-band. With other angular diameter measurement techniques, stars with dim magnitudes are difficult to observe and are necessary in determining model-independent values for stars' effective temperature and radii \citep{2013ApJ_BoyajianCHARA_MainSeqAFGKstars}. By measuring stars farther away from us and at dimmer magnitudes, we start measuring stars that are probing the edges of stellar models and their accuracy in predicting sizes of stars. In particular, cool low-mass stars are of interest in astronomy because of the difference between their expected sizes and their directly measured sizes \citep{2013ApJ_Spadaetal_radiusinflation_singlevsbinary}. 

\section{Occultation program overview}

\subsection{Hardware}
In order to observe the fringes around occultation shadows, IACTs need an optical data acquisition system (DACQ) capable of sampling rates faster than 300 Hz.
VERITAS uses an off-the-shelf DATAQ DI-710-ELS, the VERITAS enhanced current monitor (ECM). The ECM is hooked up to either 2 or 4 photomultiplier (PMT) pixels on each of the four VERITAS cameras with a maximum sampling rate of 4800 Hz. On each camera, those pixels include the central pixel and at least one background pixel to reject spurious terrestrial events. 

\begin{figure}
    \centering
    \includegraphics[width=0.95\linewidth]{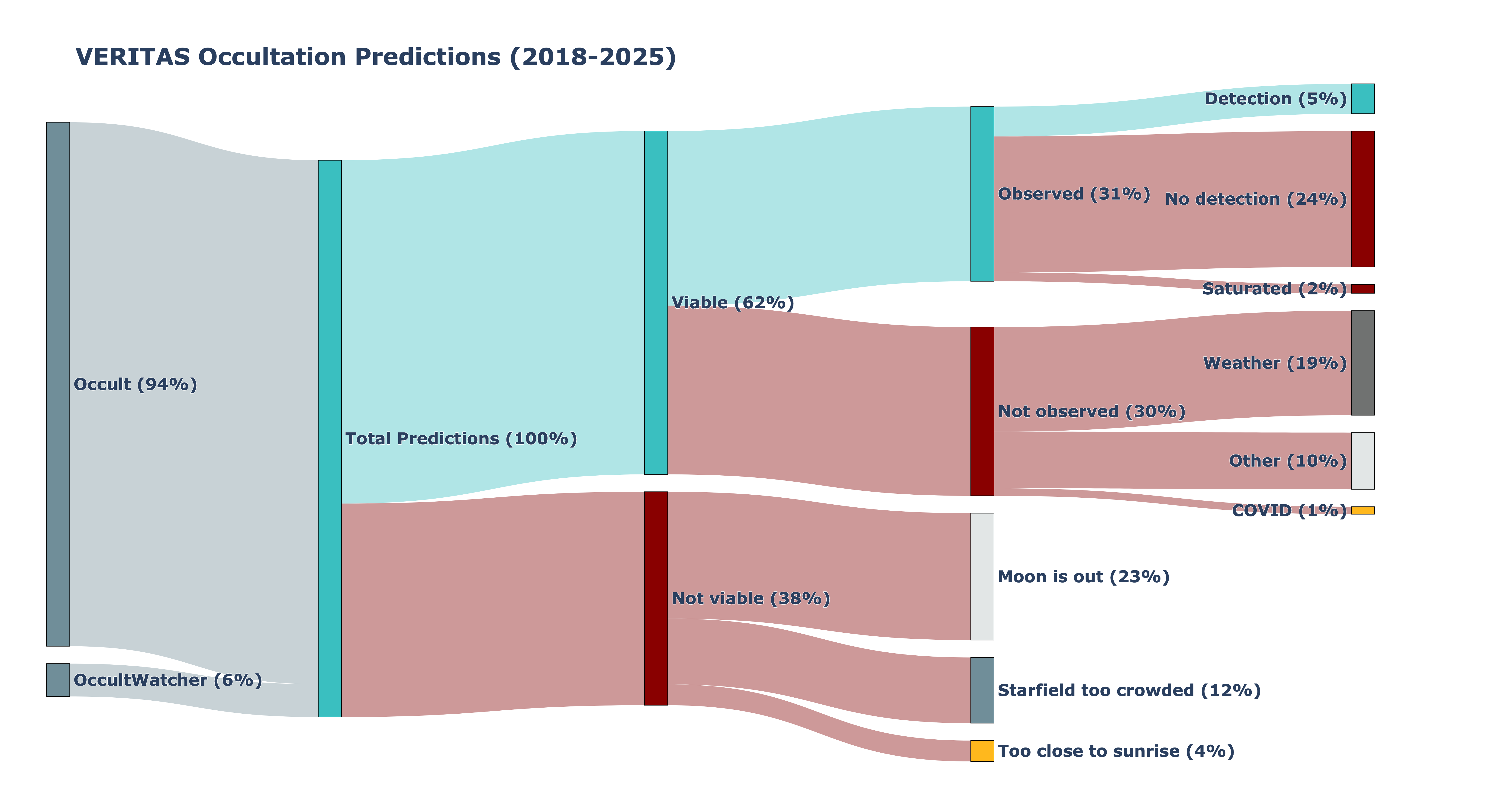}
    \caption{Occultation predictions since the 2018/1019 VERITAS observing season with their path to successful detections (shown in teal) and various other outcomes. All percentages are with respect to the total number of predictions. Outcomes other than detections are shown with red bands. Numbers for each outcome are shown along with the reason for the outcome.}
    \label{fig:Sankey_Occultationpredictions}
\end{figure}

\subsection{Occultation predictions and observations}
There are two main tools for predicting asteroid occultations: Occult\footnote{http://lunar-occultations.com/iota/occult4.htm} and OccultWatcher (OW)\footnote{https://www.occultwatcher.net/}. Occult is an active prediction software developed by David Herald that needs to run every month with 3 different ephemeris databases. OW is a passive software that regularly updates its list of predictions based on internal calculations as well as user-added predictions. Predictions from both tools are based on the observer's location and are accurate for up to 1 month. 

When making predictions for VERITAS, we filter the predictions to only consider IACT-feasible occultations (within dark time, above 20 degrees elevation, m$_{V}<13.5$, and moonlight $<66\%$ illuminated). After those filters, we only select occultations longer than 0.5 seconds in duration with a predicted probability $>10\%$ (Fig. \ref{fig:Sankey_Occultationpredictions}: Not viable 38\%).
For each prediction that passes our filters, we attempt an observation. 
Each observation is scheduled to begin 3 minutes before the time of occultation and last 5 minutes. 
Sometimes scheduled observations are canceled due to weather, hardware issues, and observations of higher priority targets (Fig. \ref{fig:Sankey_Occultationpredictions}: Not observed 30\%).

\begin{figure}
    \centering
    \includegraphics[width=0.99\linewidth]{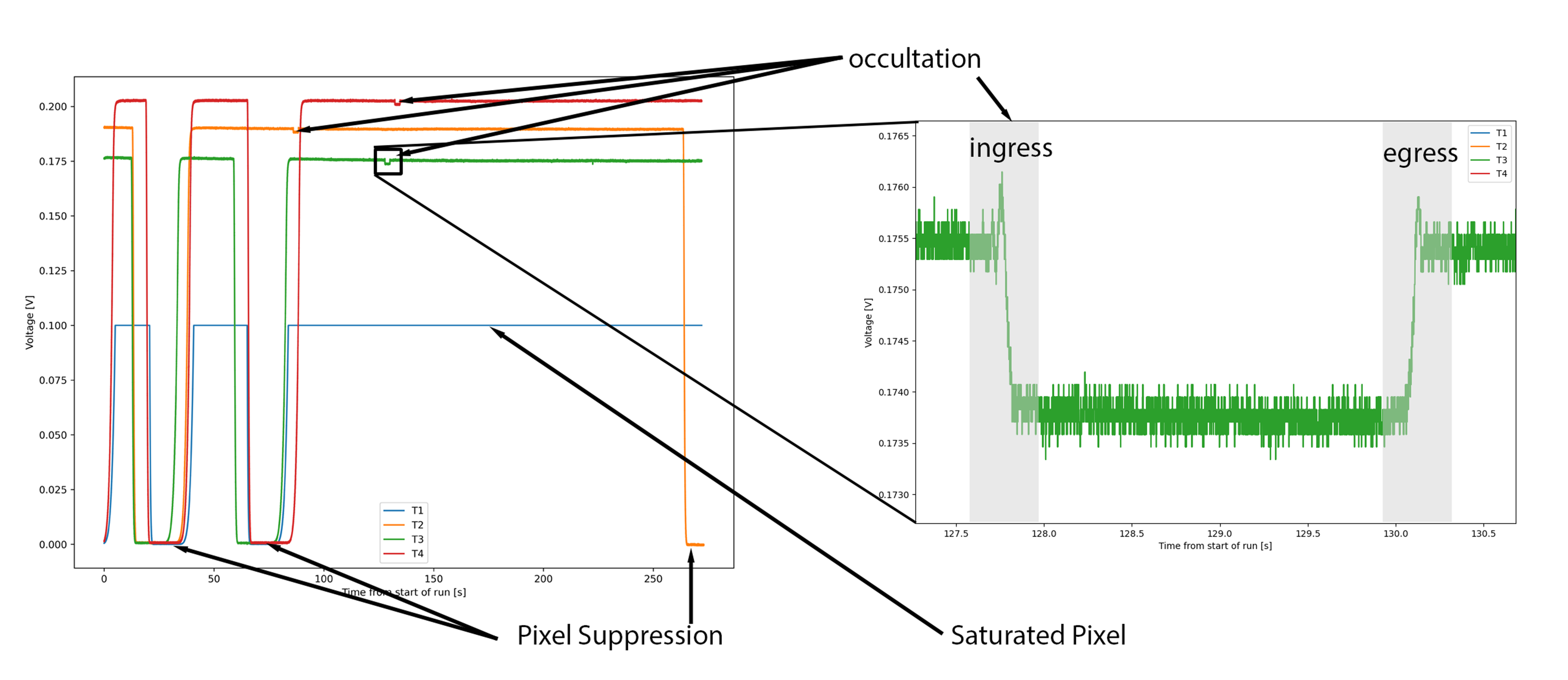}
    \caption{Left: Occultation data observed with the central pixel on each VERITAS telescope ECM with two fully visible pixel suppressions, a saturated pixel in T1, and the occultation. Left: zoomed in plot of the T3 lightcurve around the occultation with visible diffraction.}
    \label{fig:data_occultation_with_commentary}
\end{figure}

Even when observations are successfully scheduled and observed, there are other issues that can arise. The first issue is that, due to uncertainties in an asteroid's orbital parameters or a star's position, the occultation shadow does not fall over VERITAS, resulting in non-detections (Fig. \ref{fig:Sankey_Occultationpredictions}: No detection 24\%). In some cases, weather can lead to non-detections as well. The third issue to arise during observations is a saturated pixel due to
an ECM gain set too high for the stellar field resulting in a saturated pixel (Fig. \ref{fig:Sankey_Occultationpredictions}: Saturated 2\%). See Figure \ref{fig:data_occultation_with_commentary} to see saturation at play in an observation with an occultation.

\section{Occultation program outcomes}

Since 2019, VERITAS has predicted more than 300 occultations with a probability greater than 10\%. VERITAS attempted to observe 117 of these predictions. Of the 117 observations, 20 successfully observed the occultation, with 11 of the observed occultations showing at least one diffraction peak.
In total, VERITAS has observed 585 minutes of occultations, which averages to ~30 minutes of observation time per detected occultation.

\section{Occultation program considerations}
After seven seasons at VERITAS of predicting, observing, and analyzing asteroid occultations, here are some considerations for future IACT occultation observations.
\subsection{Predictions}
In making occultation predictions, there are a couple of improvements worth mentioning. First, the OW prediction software has a much higher detection rate ($\frac{7}{22}=32\%$) than Occult ($\frac{13}{351}=3.7\%$) and requires less investment of time. 
Second, getting accurate asteroid occultation orbital information relevant for the observer is very important for the angular diameter analysis. Originally, we used the JPL Horizons web ephemeris generator's "VmagOb" for obtaining the velocity of the shadow on the ground. "VmagOb" is the velocity vector of the object with respect to the observer, however this is not the same as the projected speed on the ground. This difference in definition means that "VmagOb" has been off by more than 7 km/s for some published
occultations. 
We now use the stellar occultation reduction and analysis software (SORA) \citep{2022MNRAS_SORAcitation} to calculate accurate post-observation asteroid occultation parameters such as the distance to the asteroid and the velocity of the asteroid shadow on the ground.

\subsection{Optical bandwidth of an IACT}
For an IACT, the optical bandwidth of observable wavelengths is the main limiting factor in observing occultations, since the relative heights of the peak fringes decrease with increasing bandwidth. VERITAS uses a Hamamatsu R10560-100-20 PMT with QE peaked in the blue/UV \citep{2011ICRC_Nepomuk_PMT_QEtest}. The full wavelength profile of VERITAS, including the measured mirror reflectivity, has a FWHM bandwidth of ~146 nanometers (nm) centered at 435 nm. With VERITAS' bandwidth, the limiting angular resolution is ~0.1 mas for occultations of asteroids at 3.2 AU \citep{1977AJ_Ridgway_angularlimitations_LO} (See Fig. \ref{fig:occultationEdge_DiffractionPatterns}). At the distance of TNOs 
\textasciitilde45 AU, the limiting angular resolution is \textasciitilde{0.03 mas}.
To detect stellar angular diameters below this threshold, both the instrument’s wavelength response function (IWRF) and the stellar spectrum must be known with high precision. Although the number of significant peaks will always be limited due to the bandwidth, the change in the visible peaks' heights is directly connected to the angular diameter of the occulted star (see Figures \ref{fig:occultationEdge_DiffractionPatterns}).
\subsection{Sampling rate and aperture}
While bandwidth leads to a very large impact on observed diffraction patterns, sampling rate has very little impact on the observed fringes at rates faster than 300 Hz.  Aperture also has very little impact on the observed diffraction pattern unless one observes close asteroid occultations ($\leq1.5$ AU) with very large apertures (23.0 meters).
\subsection{Timing resolution}
There are two considerations on the topic of timing: absolute timing and time delay. Additional scientific outputs of asteroid occultations are measurements of asteroid diameters, shapes, and astrometry. With precise timing information, we can fit the observed shadow durations from multiple telescopes and then calculate the length of each chord across the asteroid. However, the data collection software only provides timestamps on decimal level timescales, which leaves us with a maximum accuracy on the timescale of 0.1 seconds, enough to provide basic constraints on asteroid diameters and shapes, but not enough to add accurate astrometry data for the asteroid.
Second, in VERITAS' current monitor backend, the circuit had an RC time constant of \textasciitilde3 milliseconds, which showed up in fast optical transients in the ECM. To improve the system in November 2024, we replaced the capacitor on the circuit with a lower capacitance and now have a \textasciitilde1.5 ms time constant. 
This has a minimal impact on current asteroid occultations but may affect short-duration occultations.

\subsection{Optics}
The VERITAS optical system has a field of view of 3.5 degrees and 499 PMT pixels. Each PMT is 0.167 degrees across and the integrated light within that field of view from bright stars and the overall night sky background (NSB) can cause variations on the currents observed depending on the star field. Smaller pixels would lead to better isolation of the target star and reduce the integrated NSB light. Current IACTs are limited by their optical point spread function, which is about the same size as their pixels, but future generation IACTs like the 8-m Schwarzschild-Couder Telescope will have much smaller pixel sizes.

\subsection{Shape of asteroids}
\begin{figure}
    \centering
    \includegraphics[width=0.6\linewidth]{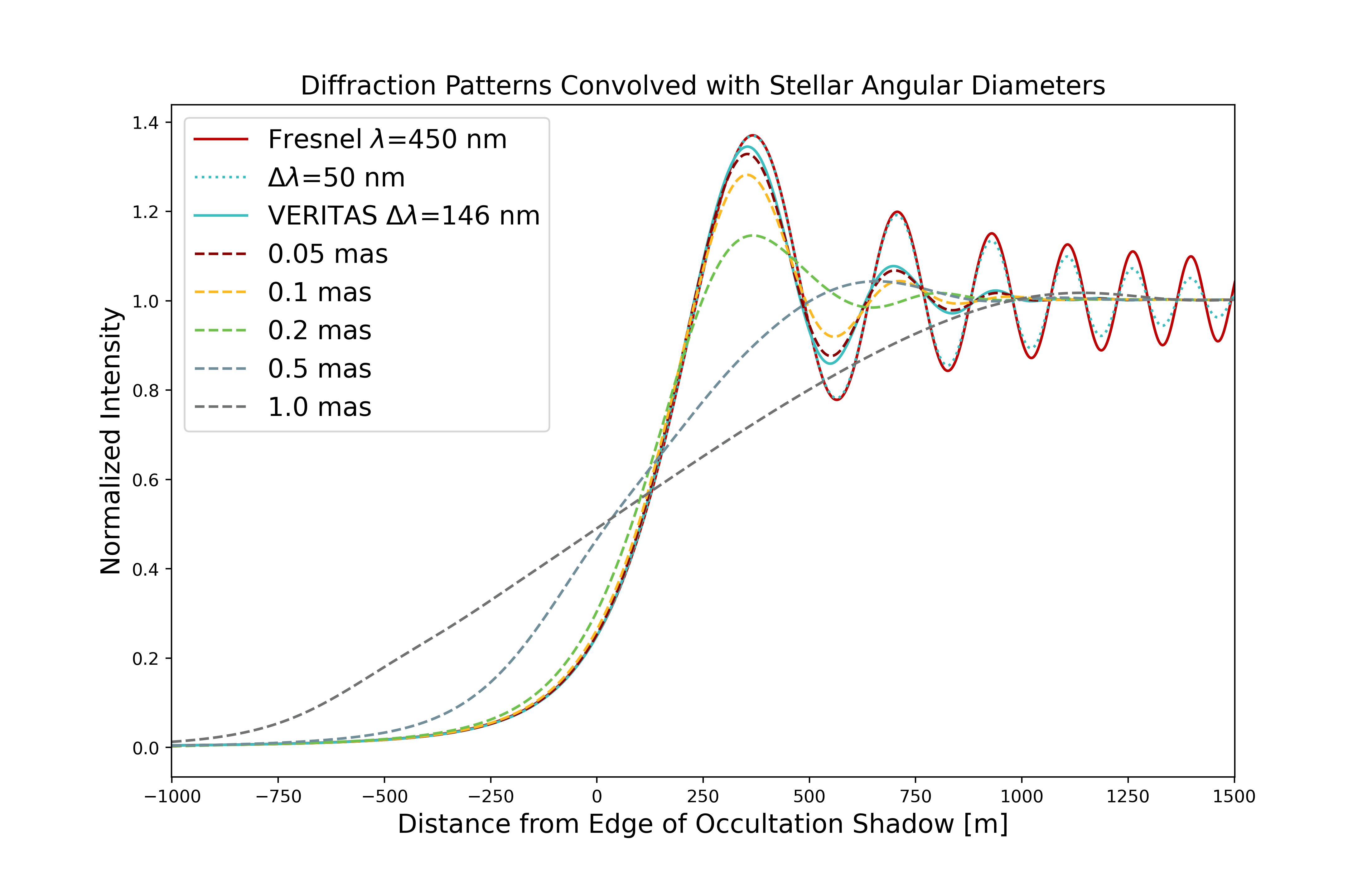}
    \caption{Intensity vs. distance from the edge of the occultation shadow. Displayed here are the monochromatic Fresnel diffraction curve, a 50-nm bandpass filtered diffraction curve, the VERITAS instrument's wavelength response function (IRWF)-corrected diffraction curve, and the diffraction curves for various stellar angular diameters with sizes in milliarcseconds convolved with the VERITAS IWRF-corrected diffraction curve.}
    \label{fig:occultationEdge_DiffractionPatterns}
\end{figure}

Another consideration that is important in the analysis of asteroid occultations is the shape of each asteroid. Because every occultation has both a covering (ingress) and uncovering (egress), there are always two diffraction patterns per telescope with which to measure the stellar angular diameter. However, the angular diameter measurement can be different between the ingress and egress, even after taking into account all systematics of the instrument and the stellar spectrum.
This discrepancy warrants further investigation to understand its origin, however it is likely due to the non-uniform shape of asteroids that can cause a reduction of the fringes (as seen in Figure \ref{fig:occultationEdge_DiffractionPatterns} similar to the impact of a larger star. Thus, when applicable, we always take the smaller of the ingress/egress measurements as the more accurate measured angular diameter of the star \citep{2021MNRAS_DyachenkoRichichi_AsO}. 

\subsection{Observing time}
After the successful detection of two asteroid occultations in 2018 \citep{2019NatAs_BenbowVERITAS_AsO}, VERITAS included asteroid occultations in its long-term observing plan, with 5 hours of dedicated observing time each year. 
However, the time spent observing individual occultations is small. 
Occultations typically range in duration from less than a second to tens of seconds, and observations are usually scheduled for only five minutes. The five-minute window ensures that gains can be properly set in case of saturation and timing calibration can be performed within the run. With an integrated GPS clock, the observation time can be reduced to less than 5 minutes per occultation. Longer observations may be necessary to achieve further science goals, such as detecting asteroid satellites and atmospheres or rings around small solar system bodies. The longer duration observations (up to 10 minutes) are necessary when measuring relatively large separations between asteroids and their satellites.

\section{Conclusions and future outlook}
The main takeaways from the last seven years of running an occultation program have been: 1) We improved the occultation software used for making predictions (OW>Occult). 2) We have found more accurate methods to determine accurate asteroid parameters (SORA) \citep{2022MNRAS_SORAcitation}. 3) We learned that the main limit of our angular resolution is the bandpass of VERITAS. 4) We improved the time delay by a factor of 2 for fast optical transients. 5) The shape of asteroids can enlarge the angular size of the star within our analysis.

By observing asteroid occultations, IACTs have much to gain with minimal overhead to observing time and cost. 
IACTs can significantly contribute to the science of stellar angular diameters at scales no current optical telescopes are capable of achieving. This is similar in scope to stellar intensity interferometry but has a much lower cost and time consideration with similar scientific output in studying stellar angular diameters. By equipping IACTs with kHz current readout, IACTs can be transformed into a premier optical telescope for measuring the smallest angular diameters of any optical observatory \citep{2019NatAs_BenbowVERITAS_AsO}.

There are also benefits to the broader impacts of IACTs. One of which is connection with local astronomical communities. The International Occultation Timing Association (IOTA) is a large multinational network of amateur astronomers who observe occultations \citep{2020MNRAS_Herald_IOTA_AsO}. By attempting to observe asteroid occultations alongside IOTA observers, IACT observatories can engage in citizen science and public outreach simultaneously.

Another broader impact of looking at asteroid occultations is in the field of solar system astronomy. Asteroid occultations and other similar stellar occultations by small bodies are used by the planetary science community to determine the sizes and shapes of objects. With large collecting areas and precise time sampling, IACTs would be unmatched at resolving the finer details of small bodies in the solar system. Other studies of small bodies using occultations have discovered rings around Centaurs and TNOs, satellites of asteroids, contact binaries, and binary star systems \citep{2017AJ_Leiva_CharikloSizeShape_Occ}. IACTs have the potential to contribute significantly to the planetary science community.

Planetary scientists are also interested in studying the properties of TNOs in order to understand the formation of the Solar System \citep{2025RSPTA_Buie_SolarSysFormation_Occ}. 
Targeted observations of TNO occultations have led to the discovery of atmospheres around TNOs as well as rings, satellites, and binary TNO systems \citep{2017AJ_Leiva_CharikloSizeShape_Occ}. 
In the future, IACTs can contribute to these targeted observations and provide high angular resolution to studies about the surfaces of TNOs. 
Additionally, by equipping IACTs with ECM-like data acquisition systems across their entire field of view we can search for serendipitous occultations of small sub-km TNOs and set constraints on the formation of the solar system \citep{2012ApJ_Schlichtingetal_HST_serendipitousoccultationsearch}. This idea for upgrading the entire field of view of an IACT for optical observations is currently being developed at VERITAS.

\acknowledgments 
This research is supported by grants from the U.S. Department of Energy, the
U.S. National Science Foundation and the Smithsonian Institution, by NSERC in Canada, by
the Helmholtz Association in Germany, and by STFC in the UK. 
We acknowledge the excellent work of the technical support staff at
the Fred Lawrence Whipple Observatory and at the collaborating institutions in the construction and operation of the instrument.
J. Bartkoske gratefully acknowledges support under NSF award \texttt{\#}2411860.

\bibliographystyle{JHEP}
\bibliography{AsteroidOccultationPapers}

@ARTICLE{2019NatAs_BenbowVERITAS_AsO,
       author = {{Benbow}, W. and {Bird}, R. and {Brill}, A. and {Brose}, R. and {Chromey}, A.~J. and {Daniel}, M.~K. and {Feng}, Q. and {Finley}, J.~P. and {Fortson}, L. and {Furniss}, A. and {Gillanders}, G.~H. and {Giuri}, C. and {Gueta}, O. and {Hanna}, D. and {Halpern}, J.~P. and {Hassan}, T. and {Holder}, J. and {Hughes}, G. and {Humensky}, T.~B. and {Joyce}, A.~M. and {Kaaret}, P. and {Kar}, P. and {Kelley-Hoskins}, N. and {Kertzman}, M. and {Kieda}, D. and {Krause}, M. and {Lang}, M.~J. and {Lin}, T.~T.~Y. and {Maier}, G. and {Matthews}, N. and {Moriarty}, P. and {Mukherjee}, R. and {Nieto}, D. and {Nievas-Rosillo}, M. and {O'Brien}, S. and {Ong}, R.~A. and {Park}, N. and {Petrashyk}, A. and {Pohl}, M. and {Pueschel}, E. and {Quinn}, J. and {Ragan}, K. and {Reynolds}, P.~T. and {Richards}, G.~T. and {Roache}, E. and {Rulten}, C. and {Sadeh}, I. and {Santander}, M. and {Sembroski}, G.~H. and {Shahinyan}, K. and {Sushch}, I. and {Wakely}, S.~P. and {Wells}, R.~M. and {Wilcox}, P. and {Wilhelm}, A. and {Williams}, D.~A. and {Williamson}, T.~J.},
        title = "{Direct measurement of stellar angular diameters by the VERITAS Cherenkov telescopes}",
      journal = {Nature Astronomy},
         year = 2019,
        month = apr,
       volume = {3},
        pages = {511-516},
          doi = {10.1038/s41550-019-0741-z},
archivePrefix = {arXiv},
       eprint = {1904.06324},
 primaryClass = {astro-ph.SR},
       adsurl = {https://ui.adsabs.harvard.edu/abs/2019NatAs...3..511B}
}

@ARTICLE{2013ApJ_BoyajianCHARA_MainSeqAFGKstars,
       author = {{Boyajian}, Tabetha S. and {von Braun}, Kaspar and {van Belle}, Gerard and {Farrington}, Chris and {Schaefer}, Gail and {Jones}, Jeremy and {White}, Russel and {McAlister}, Harold A. and {ten Brummelaar}, Theo A. and {Ridgway}, Stephen and {Gies}, Douglas and {Sturmann}, Laszlo and {Sturmann}, Judit and {Turner}, Nils H. and {Goldfinger}, P.~J. and {Vargas}, Norm},
        title = "{Stellar Diameters and Temperatures. III. Main-sequence A, F, G, and K Stars: Additional High-precision Measurements and Empirical Relations}",
      journal = {\apj},
         year = 2013,
        month = jul,
       volume = {771},
       number = {1},
          eid = {40},
        pages = {40},
          doi = {10.1088/0004-637X/771/1/40},
archivePrefix = {arXiv},
       eprint = {1306.2974},
 primaryClass = {astro-ph.SR},
       adsurl = {https://ui.adsabs.harvard.edu/abs/2013ApJ...771...40B}
}

@article{1936AcaDeSci_Arnulf_firstLOmeasurement,
       author = {Albert Arnulf},
        title = {Sur une m\'ethode pour la mesure des diam\'etres apparents des \'etoiles},
      journal = {Comptes Rendus de l'Acad\'emie des Sciences},
         year = 1936,
        month = jan,
       volume = {202},
        pages = {115--117}
}

@ARTICLE{2021MNRAS_DyachenkoRichichi_AsO,
       author = {{Dyachenko}, V. and {Richichi}, A. and {Obolentseva}, M. and {Beskakotov}, A. and {Maksimov}, A. and {Mitrofanova}, A. and {Balega}, Yu},
        title = "{A joint occultation and speckle investigation of the binary star TYC 1947-290-1 and of the asteroid (87) Sylvia}",
      journal = {\mnras},
         year = 2021,
        month = dec,
       volume = {508},
       number = {2},
        pages = {2730-2735},
          doi = {10.1093/mnras/stab2767},
       adsurl = {https://ui.adsabs.harvard.edu/abs/2021MNRAS.508.2730D}
}

@ARTICLE{2020MNRAS_Herald_IOTA_AsO,
       author = {{Herald}, David and {Gault}, David and {Anderson}, Robert and {Dunham}, David and {Frappa}, Eric and {Hayamizu}, Tsutomu and {Kerr}, Steve and {Miyashita}, Kazuhisa and {Moore}, John and {Pavlov}, Hristo and {Preston}, Steve and {Talbot}, John and {Timerson}, Brad},
        title = "{Precise astrometry and diameters of asteroids from occultations - a data set of observations and their interpretation}",
      journal = {\mnras},
         year = 2020,
        month = dec,
       volume = {499},
       number = {3},
        pages = {4570-4590},
          doi = {10.1093/mnras/staa3077},
archivePrefix = {arXiv},
       eprint = {2010.06086},
 primaryClass = {astro-ph.EP},
       adsurl = {https://ui.adsabs.harvard.edu/abs/2020MNRAS.499.4570H}
}

@ARTICLE{2017AJ_Leiva_CharikloSizeShape_Occ,
       author = {{Leiva}, R. and {Sicardy}, B. and {Camargo}, J.~I.~B. and {Ortiz}, J. -L. and {Desmars}, J. and {B{\'e}rard}, D. and {Lellouch}, E. and {Meza}, E. and {Kervella}, P. and {Snodgrass}, C. and {Duffard}, R. and {Morales}, N. and {Gomes-J{\'u}nior}, A.~R. and {Benedetti-Rossi}, G. and {Vieira-Martins}, R. and {Braga-Ribas}, F. and {Assafin}, M. and {Morgado}, B.~E. and {Colas}, F. and {De Witt}, C. and {Sickafoose}, A.~A. and {Breytenbach}, H. and {Dauvergne}, J. -L. and {Schoenau}, P. and {Maquet}, L. and {Bath}, K. -L. and {Bode}, H. -J. and {Cool}, A. and {Lade}, B. and {Kerr}, S. and {Herald}, D.},
        title = "{Size and Shape of Chariklo from Multi-epoch Stellar Occultations}",
      journal = {\aj},
         year = 2017,
        month = oct,
       volume = {154},
       number = {4},
          eid = {159},
        pages = {159},
          doi = {10.3847/1538-3881/aa8956},
archivePrefix = {arXiv},
       eprint = {1708.08934},
 primaryClass = {astro-ph.EP},
       adsurl = {https://ui.adsabs.harvard.edu/abs/2017AJ....154..159L}
}

@ARTICLE{2025RSPTA_Buie_SolarSysFormation_Occ,
       author = {{Buie}, Marc W. and {Keller}, John M. and {Nesvorn{\'y}}, David and {Porter}, Simon B.},
        title = "{Occultation constraints on solar system formation models}",
      journal = {Philosophical Transactions of the Royal Society of London Series A},
         year = 2025,
        month = feb,
       volume = {383},
       number = {2291},
          eid = {20240194},
        pages = {20240194},
          doi = {10.1098/rsta.2024.0194},
archivePrefix = {arXiv},
       eprint = {2502.00062},
 primaryClass = {astro-ph.EP},
       adsurl = {https://ui.adsabs.harvard.edu/abs/2025RSPTA.38340194B}
}

@ARTICLE{2013ApJ_Spadaetal_radiusinflation_singlevsbinary,
       author = {{Spada}, F. and {Demarque}, P. and {Kim}, Y. -C. and {Sills}, A.},
        title = "{The Radius Discrepancy in Low-mass Stars: Single versus Binaries}",
      journal = {\apj},
         year = 2013,
        month = oct,
       volume = {776},
       number = {2},
          eid = {87},
        pages = {87},
          doi = {10.1088/0004-637X/776/2/87},
archivePrefix = {arXiv},
       eprint = {1308.5558},
 primaryClass = {astro-ph.SR},
       adsurl = {https://ui.adsabs.harvard.edu/abs/2013ApJ...776...87S}
}

@ARTICLE{2012ApJ_Schlichtingetal_HST_serendipitousoccultationsearch,
       author = {{Schlichting}, Hilke E. and {Ofek}, Eran O. and {Sari}, Re'em and {Nelan}, Edmund P. and {Gal-Yam}, Avishay and {Wenz}, Michael and {Muirhead}, Philip and {Javanfar}, Nikta and {Livio}, Mario},
        title = "{Measuring the Abundance of Sub-kilometer-sized Kuiper Belt Objects Using Stellar Occultations}",
      journal = {\apj},
         year = 2012,
        month = dec,
       volume = {761},
       number = {2},
          eid = {150},
        pages = {150},
          doi = {10.1088/0004-637X/761/2/150},
archivePrefix = {arXiv},
       eprint = {1210.8155},
 primaryClass = {astro-ph.EP},
       adsurl = {https://ui.adsabs.harvard.edu/abs/2012ApJ...761..150S}
}

@INPROCEEDINGS{2019ICRC_HassanDaniel_AsO_VERITAS_Overview,
       author = {{Hassan}, T. and {Daniel}, M.},
        title = "{Proving the outstanding capabilities of Imaging Atmospheric Cherenkov Telescopes in high time resolution optical astronomy}",
    booktitle = {36th International Cosmic Ray Conference (ICRC2019)},
         year = 2019,
       series = {International Cosmic Ray Conference},
       volume = {36},
        month = jul,
          eid = {692},
        pages = {692},
          doi = {10.22323/1.358.0692},
archivePrefix = {arXiv},
       eprint = {1908.03393},
 primaryClass = {astro-ph.IM},
       adsurl = {https://ui.adsabs.harvard.edu/abs/2019ICRC...36..692H}
}

@INPROCEEDINGS{2011ICRC_Nepomuk_PMT_QEtest,
       author = {{OTTE}, Nepomuk},
        title = "{Upgrade of VERITAS with high efficiency photomultipliers}",
    booktitle = {International Cosmic Ray Conference},
         year = 2011,
       series = {International Cosmic Ray Conference},
       volume = {9},
        month = jan,
        pages = {247},
          doi = {10.7529/ICRC2011/V09/1305},
       adsurl = {https://ui.adsabs.harvard.edu/abs/2011ICRC....9..247O}
}

@ARTICLE{1977AJ_Ridgway_angularlimitations_LO,
       author = {{Ridgway}, S.~T.},
        title = "{Considerations for the application of the lunar occultation technique.}",
      journal = {\aj},
         year = 1977,
        month = jul,
       volume = {82},
        pages = {511-515},
          doi = {10.1086/112083},
       adsurl = {https://ui.adsabs.harvard.edu/abs/1977AJ.....82..511R}
}

@ARTICLE{2022MNRAS_SORAcitation,
       author = {{Gomes-J{\'u}nior}, A.~R. and {Morgado}, B.~E. and {Benedetti-Rossi}, G. and {Boufleur}, R.~C. and {Rommel}, F.~L. and {Banda-Huarca}, M.~V. and {Kilic}, Y. and {Braga-Ribas}, F. and {Sicardy}, B.},
        title = "{SORA: Stellar occultation reduction and analysis}",
      journal = {\mnras},
         year = 2022,
        month = mar,
       volume = {511},
       number = {1},
        pages = {1167-1181},
          doi = {10.1093/mnras/stac032},
archivePrefix = {arXiv},
       eprint = {2201.01799},
 primaryClass = {astro-ph.IM},
       adsurl = {https://ui.adsabs.harvard.edu/abs/2022MNRAS.511.1167G}
}
%

\section*{Full Author List: VERITAS Collaboration}

\scriptsize
\noindent
A.~Archer$^{1}$,
P.~Bangale$^{2}$,
J.~T.~Bartkoske$^{3}$,
W.~Benbow$^{4}$,
Y.~Chen$^{5}$,
J.~L.~Christiansen$^{6}$,
A.~J.~Chromey$^{4}$,
A.~Duerr$^{3}$,
M.~Errando$^{7}$,
M.~Escobar~Godoy$^{8}$,
J.~Escudero Pedrosa$^{4}$,
Q.~Feng$^{3}$,
S.~Filbert$^{3}$,
L.~Fortson$^{9}$,
A.~Furniss$^{8}$,
W.~Hanlon$^{4}$,
O.~Hervet$^{8}$,
C.~E.~Hinrichs$^{4,10}$,
J.~Holder$^{11}$,
T.~B.~Humensky$^{12,13}$,
M.~Iskakova$^{7}$,
W.~Jin$^{5}$,
M.~N.~Johnson$^{8}$,
E.~Joshi$^{14}$,
M.~Kertzman$^{1}$,
M.~Kherlakian$^{15}$,
D.~Kieda$^{3}$,
T.~K.~Kleiner$^{14}$,
N.~Korzoun$^{11}$,
S.~Kumar$^{12}$,
M.~J.~Lang$^{16}$,
M.~Lundy$^{17}$,
G.~Maier$^{14}$,
C.~E~McGrath$^{18}$,
P.~Moriarty$^{16}$,
R.~Mukherjee$^{19}$,
W.~Ning$^{5}$,
R.~A.~Ong$^{5}$,
A.~Pandey$^{3}$,
M.~Pohl$^{20,14}$,
E.~Pueschel$^{15}$,
J.~Quinn$^{18}$,
P.~L.~Rabinowitz$^{7}$,
K.~Ragan$^{17}$,
P.~T.~Reynolds$^{21}$,
D.~Ribeiro$^{9}$,
E.~Roache$^{4}$,
I.~Sadeh$^{14}$,
L.~Saha$^{4}$,
H.~Salzmann$^{8}$,
M.~Santander$^{22}$,
G.~H.~Sembroski$^{23}$,
B.~Shen$^{12}$,
M.~Splettstoesser$^{8}$,
A.~K.~Talluri$^{9}$,
S.~Tandon$^{19}$,
J.~V.~Tucci$^{24}$,
J.~Valverde$^{25,13}$,
V.~V.~Vassiliev$^{5}$,
D.~A.~Williams$^{8}$,
S.~L.~Wong$^{17}$,
T.~Yoshikoshi$^{26}$\\
\\
\noindent
$^{1}${Department of Physics and Astronomy, DePauw University, Greencastle, IN 46135-0037, USA}

\noindent
$^{2}${Department of Physics, Temple University, Philadelphia, PA 19122, USA}

\noindent
$^{3}${Department of Physics and Astronomy, University of Utah, Salt Lake City, UT 84112, USA}

\noindent
$^{4}${Center for Astrophysics $|$ Harvard \& Smithsonian, Cambridge, MA 02138, USA}

\noindent
$^{5}${Department of Physics and Astronomy, University of California, Los Angeles, CA 90095, USA}

\noindent
$^{6}${Physics Department, California Polytechnic State University, San Luis Obispo, CA 94307, USA}

\noindent
$^{7}${Department of Physics, Washington University, St. Louis, MO 63130, USA}

\noindent
$^{8}${Santa Cruz Institute for Particle Physics and Department of Physics, University of California, Santa Cruz, CA 95064, USA}

\noindent
$^{9}${School of Physics and Astronomy, University of Minnesota, Minneapolis, MN 55455, USA}

\noindent
$^{10}${Department of Physics and Astronomy, Dartmouth College, 6127 Wilder Laboratory, Hanover, NH 03755 USA}

\noindent
$^{11}${Department of Physics and Astronomy and the Bartol Research Institute, University of Delaware, Newark, DE 19716, USA}

\noindent
$^{12}${Department of Physics, University of Maryland, College Park, MD, USA }

\noindent
$^{13}${NASA GSFC, Greenbelt, MD 20771, USA}

\noindent
$^{14}${DESY, Platanenallee 6, 15738 Zeuthen, Germany}

\noindent
$^{15}${Fakult\"at f\"ur Physik \& Astronomie, Ruhr-Universit\"at Bochum, D-44780 Bochum, Germany}

\noindent
$^{16}${School of Natural Sciences, University of Galway, University Road, Galway, H91 TK33, Ireland}

\noindent
$^{17}${Physics Department, McGill University, Montreal, QC H3A 2T8, Canada}

\noindent
$^{18}${School of Physics, University College Dublin, Belfield, Dublin 4, Ireland}

\noindent
$^{19}${Department of Physics and Astronomy, Barnard College, Columbia University, NY 10027, USA}

\noindent
$^{20}${Institute of Physics and Astronomy, University of Potsdam, 14476 Potsdam-Golm, Germany}

\noindent
$^{21}${Department of Physical Sciences, Munster Technological University, Bishopstown, Cork, T12 P928, Ireland}

\noindent
$^{22}${Department of Physics and Astronomy, University of Alabama, Tuscaloosa, AL 35487, USA}

\noindent
$^{23}${Department of Physics and Astronomy, Purdue University, West Lafayette, IN 47907, USA}

\noindent
$^{24}${Department of Physics, Indiana University Indianapolis, Indianapolis, Indiana 46202, USA}

\noindent
$^{25}${Department of Physics, University of Maryland, Baltimore County, Baltimore MD 21250, USA}

\noindent
$^{26}${Institute for Cosmic Ray Research, University of Tokyo, 5-1-5, Kashiwa-no-ha, Kashiwa, Chiba 277-8582, Japan}

\end{document}